# Comparison of phase change process in Si-GST hybrid integrated waveguide and MMI devices

Hanyu Zhang,[1] Xing Yang,[2] Liangjun lu,[2,*] Jianping Chen,[2] B. M. A. Rahman,[3] and Linjie Zhou[2]

[1] *College of Telecommunications and Information Engineering, Nanjing University of Posts and Telecommunications, Nanjing 210023, China*
[2] *State Key Laboratory of Advanced Optical Communication Systems and Networks, Shanghai Institute for Advanced Communication and Data Science, Department of Electronic Engineering, Shanghai Jiao Tong University, Shanghai 200240, China*
[3] *Department of Electrical and Electronic Engineering, City, University of London, London EC 1V0HB, U. K.*
*\* luliangjun@sjtu.edu.cn*

**Abstract:** In the past decades, silicon photonic integrated circuits (PICs) have been considered as a promising approach to solve the bandwidth bottleneck in optical communications and interconnections. Despite significant advances, large-scale PICs still face a series of technical challenges, such as footprint, power consumption, and routing state storage, resulting from the active tuning methods used to control the optical waves. These challenges can be partially addressed by combining chalcogenide phase change materials (PCMs) such as $Ge_2Sb_2Te_5$ (GST) with silicon photonics, especially applicable in switching applications due to the nonvolatile nature of the GST. Although GST phase transitions between amorphous and crystalline states actuated by optical and electrical pulses heating have been experimentally demonstrated, there is no direct comparison between them. We carried out simulations and experiments to systematically investigate the difference in the phase change process induced by optical and electrical pulses for two types of Si-GST hybrid waveguides. For the phase transition induced by optical pulses, the device has a clear advantage in terms of power consumption and operation speed. For the phase transition induced by electrical pulses, the device is suitable for large-scale integration because it does not require complex light routing. It helps us better understand the phase change process and push forward the further development of Si-GST hybrid integration platform, bringing in new potential applications.

## 1. Introduction

Silicon photonics has been widely considered as one of the most promising candidates for low-cost photonic integration by leveraging the complementary metal-oxide-semiconductor (CMOS) infrastructure [1]. In recent years, a wide range of silicon photonic devices have been demonstrated on the silicon-on-insulator (SOI) platform [2-4]. However, the active tuning methods are usually based on two mechanisms: the thermo-optic effect [5, 6] or the free-carrier dispersion effect [7-9]. It is still challenging to make tunable optical devices with a small footprint and low power consumption. The hybrid integration of new materials onto silicon for emerging applications has attracted considerable attention [10-16]. Among the various existing materials, chalcogenide-based phase change materials (PCMs) hybrid-integrated with silicon (Si) or silicon nitride ($Si_3N_4$) offers a promising solution for specific applications, such as neuromorphic computing [17-20], optical switching [21-24], all-optical memory [25, 26], and highly reconfigurable general-purpose photonic circuits resembling the field-programmable gate array (FPGA) in electronics [27-29]. $Ge_2Sb_2Te_5$ (GST) is the most commonly used PCM and has been commercially used in both optical and electronic data storage [30, 31]. It possesses at least two phase states, amorphous state (aGST) and crystalline state (cGST). The introduction of GST to silicon photonics can bring various advantages. First, GST is a nonvolatile material, which means the phase state can be maintained without any static power consumption. Therefore, the optical routing state can be stored after tuning. Second, the refractive index contrast between the amorphous and crystalline states is very large, especially in the near-infrared telecom wavelength range. It indicates that the active devices can be made very compact. Third, reversible switching between two phase states can be realized for multiple cycles at high speed. It ensures long-term operation of the devices without failure. Finally, GST can be easily deposited using radio frequency (RF) sputtering from a stoichiometric GST alloy target. It hence allows for low-cost fabrication of Si-GST hybrid devices.

The phase change process is dependent on the duration time of the excitation pulse. The underlying mechanism is related to both thermal [32] and athermal effects [33]. The thermal phase transition mechanism of GST is due to the localized heating in the presence of an optical or electrical pulse. A low-power long-duration pulse heats up the material above the glass transition temperature, inducing the phase change from the amorphous to the crystalline state. On the other hand, to reversely transform this material, a shorter but higher power pulse is necessary to generate enough heat to melt the crystalline material. After melting, the material is then cooled down by a fast thermal quenching process so that it eventually becomes amorphous. The athermal effects can only be observed when the excitation pulse is below the picosecond range. During the crystalline to the amorphous transition, the partial covalent bonds of Ge atoms are broken by the excitation pulse to induce an atomic structure migration from an octahedral to a tetrahedral configuration. The change could be realized just by electronic excitation without the need for melting [34]. Therefore, the amorphization process can occur in only several picoseconds.

In last five years, a lot of work has also been done on developing phase-change photonics using Si-GST hybrid waveguides, where optical or electrical pulses are often used to induce the phase change. The basic waveguide structures are strip waveguide [35-38] and multimode interfere (MMI) waveguide [39-42]. However, the differences in the phase transition process under optical and electrical pulses excitation for these two types of structures have not been fully investigated. In this article, we compare their basic optical properties and dynamic responses under optical and electrical pulses. Our experimental results and analyses serve to provide a reference for new device design based on the hybrid integration of silicon and GST.

## 2. Results and discussions

### 2.1 Basic optical properties

The influence of GST cell dimensions, namely, length and width, has been computationally studied for the amorphization and crystallization processes in $Si_3N_4$-GST hybrid waveguides[43]. The GST cell thickness is usually fixed at 10-20 nm. A thicker GST cell in principle gives a higher extinction coefficient and larger effective refractive index difference between amorphous and crystalline states. In this section, we investigate the influence of GST layer thickness on the insertion loss and extinction ratio for silicon strip and MMI waveguides. It should be noted that we choose silicon instead of $Si_3N_4$ waveguides because it is beneficial to control the phase state of GST by using electrical pulses, which will be discussed in Section 2.3.

Figure 1(a) shows the structures of silicon strip and MMI waveguides covered with a 1-μm-long GST cell. An indium-tin-oxide (ITO) layer with the same thickness as GST is put on top of the GST in order to prevent its oxidation in air. The MMI waveguide is based on the self-imaging principle. The input light spot is reproduced in periodic intervals in the MMI region. When the GST cell is positioned at the light focusing spot, it has substantial overlap with the optical wave, resulting in a high tuning efficiency. We first studied the basic optical properties of both devices through numerical simulations. The operation wavelength was chosen at $\lambda = 1.55$ μm. The refractive indices of the materials used in the simulations were adopted from [44]. Figure 1(b) shows the magnitude of the electrical field intensity ($|E|^2$) distributions in the *x-z* plane for the silicon strip waveguide with 30-nm-thick GST when the GST is amorphous (fig. 1(b) left) and crystalline (fig. 1(b) right). The fundamental transverse electric (TE) mode of the silicon waveguide is launched into the input waveguide. As a comparison, Figure 1(c) shows the results for the MMI waveguide with the same thickness of GST. A significant reduction in optical transmission is observed for the MMI waveguide when GST returns to crystalline state (fig. 2(b) left).

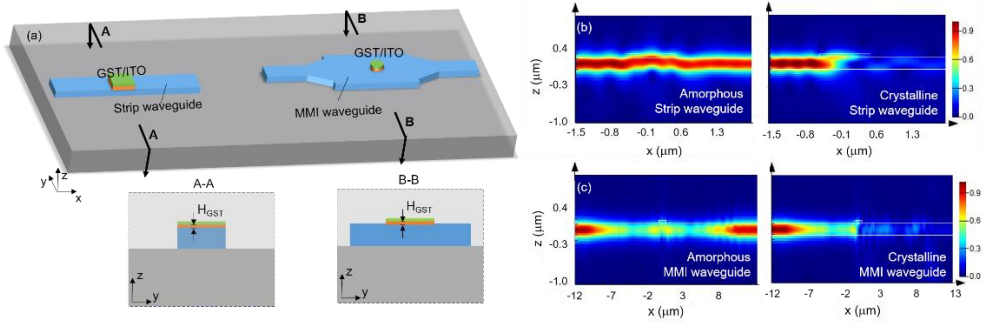

Fig. 1. (a) Schematic structures of a strip waveguide (left) and a MMI waveguide (right) with a small GST cell on top. Insets show the cross-section of the hybrid section. (b, c) Simulated electrical field intensity distributions in (b) the strip waveguide and (c) the MMI waveguide when the GST is amorphous (left) and crystalline (right).

Figure 2(a) shows the simulated transmission spectra of the strip and the MMI waveguides. We can observe that the MMI waveguide has a lower loss in the amorphous state and a higher transmission contrast between the crystalline and amorphous states. We also performed 3D finite-difference time-domain (FDTD) simulations to obtain reflection, transmission, and absorption power with different thicknesses of GST for amorphous and crystalline states, as shown in Figs. 2(b) and 2(c), respectively. The absorption loss was calculated by subtracting the transmission and reflection power from the input power. With a thicker GST, the absorption losses of the two waveguides increase in both states and the optical transmission contrast between the two states also increases. However, a thicker GST layer also gives a higher optical attenuation in the amorphous state. Figure 1(d) shows the extinction ratio of the two waveguides. The MMI waveguide exhibits a higher extinction ratio when the GST cell is thicker than 27.5 nm.

*2.2 Dynamic responses to optical pulses*

The optical transmission simulations only give the steady-state optical field distribution and leave us no information about the dynamic process during the phase change process. In the above simulations, we assumed that the GST reaches a fully crystalline or amorphous state. In this section, we study the dynamic phase change process for the two types of waveguides in the presence of an optical pump pulse. The pump pulse comes from the same waveguide and evanescently interacts with GST. We developed a comprehensive 3D finite element method (FEM) model to study the optical-thermal interactions in both waveguides. The power consumption and operation speed were compared in particular. In the simulation, an Electromagnetic Waves Beam Envelopes model from COMSOL was used to calculate the power absorbed in the GST and ITO, while a Heat Transfer in Solids model was employed to calculate the temperature distribution. Table 1 summarizes the related material properties. The refractive index of GST in both states linearly changes with temperature ($T$). The resistivity and thermal conductivity of GST have a nonlinear relationship with temperature [45].

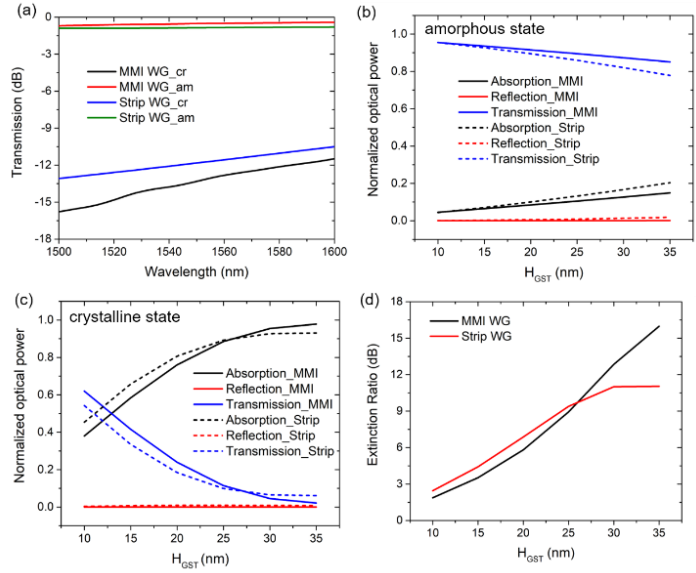

Fig. 2. (a) Simulated transmission spectra of the strip and MMI waveguides with amorphous and crystalline GST. (b, c) Normalized optical transmission, reflection, and absorption power in the Si-GST hybrid waveguide when GST is (b) amorphous and (c) crystalline. (d) Extinction Ratio between amorphous and crystalline states as a function of GST thickness.

**Table 1: Basic material properties**

| Material | Si | SiO$_2$ | ITO | aGST | cGST |
|---|---|---|---|---|---|
| Heat capacity (J/(kg·K)) | 720 | 740 | 340 | 210 | 210 |
| Thermal conductivity (W/(m·K)) | 149 | 1.38 | 11 | Variable [a] | Variable [a] |
| Density (kg/m$^3$) | 2330 | 2200 | 7100 | 6150 | 6150 |
| Complex refractive index | 3.48 | 1.45 | 1.50+0.20i | n=6.11-2.20×10$^{-4}$T k=0.83+1.56×10$^{-3}$T | n=3.94+1.11×10$^{-3}$T k=0.045+4.1×10$^{-4}$T |

[a] The resistivity and thermal conductivity of GST change with temperature [45].

We first compare the power consumption during the amorphization process. The GST layer thickness is 30 nm. Since the amorphization temperature (~600°C) is much higher than the crystalline temperature (~160°C), GST material is more easily to degrade during the amorphization process. Amorphization can be induced by a single pulse, but crystallization needs several pulses to reach a fully crystalline state. Therefore, we only simulate the amorphization process. Figures 3(a) and 3(d) show the temperature change in response to an excitation optical pulse with different power with duration of 20 ns. The inset presents the 2D temperature distribution in the lateral cross section of the GST layer at the falling edge of the optical pulse. For higher optical power, a larger amorphized GST zone can be obtained. Figures 3(b) and 3(e) show the temperature variation along the ridge-lines (light propagation direction) denoted in Figs. 3(a) and 3(d), respectively. It can be seen that temperature distribution is not uniform. The maximum-to-minimum temperature ratio is close to two for both waveguides. It suggests that the central part of GST will be first amorphized and it will easily destroy the central part if one attempts to amorphize the entire GST cell. Figure 3(c) shows the maximum temperature change as a function of pump pulse power. The stripe waveguide has lower power consumption. With an identical optical pump pulse, the GST temperature for the strip waveguide is around 2.5 times that for the MMI waveguide. Because silicon has higher thermal conductivity than silicon dioxide and air, the MMI section offers good thermal leakage, leading to a lower temperature. The temperature rises linearly with increased pump pulse power. By controlling the pump pulse power, we can get multiple intermediate states with a mixture of crystalline and amorphous GST. This characteristic increases the waveguide tuning freedom that can be employed to perform multiplication and logic computation [46]. Figure 3(f) compares the transient temperature responses for the two waveguides with different powers when the two devices are raised to the same temperature. The structure has an apparent effect on the response time. The temperature of the GST in the MMI waveguide rises faster probably because the light is focused in the MMI waveguide. Meanwhile, it also cools down faster because the wide MMI section provides a faster thermal dissipation rate. The operation speed is determined by the dead time, which is defined as the $1/e$ cooling time.

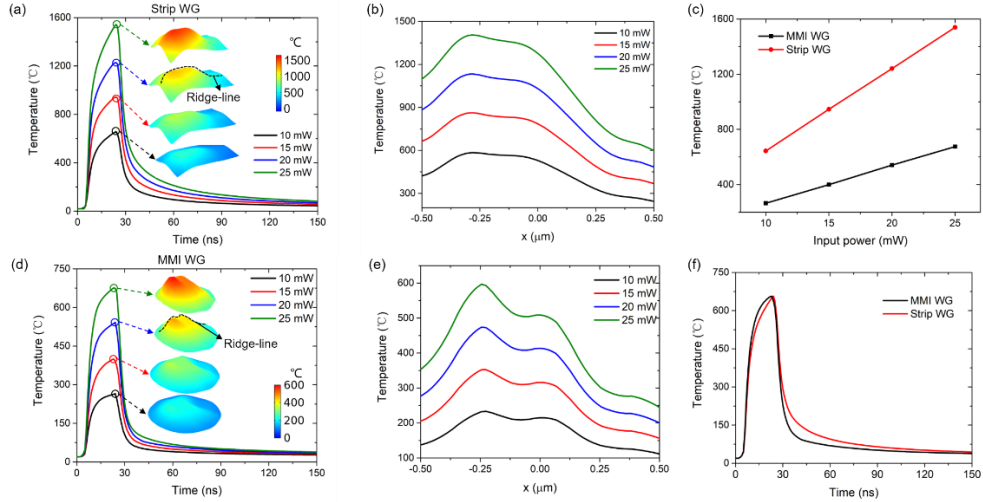

Fig. 3. (a, d) GST temperature change in response to an excitation optical pulse of different power for (a) strip waveguide (d) and MMI waveguide, respectively. Insets: temperature profiles in the planar cross-section of the GST layer. (b, e) Temperature distribution along the ridge-lines for (b) strip waveguide (e) and MMI waveguide. (c) The maximum temperature in GST change as a function of pump pulse energy. (f) The temporal temperature responses when the two waveguides are raised to the same temperature.

GST layer thickness also significantly affects power consumption and operation speed. Figure 4(a) shows the comparison of temperature change with different thicknesses of GST. The amorphization process uses a single pulse with a 20 ns duration and 10 mW peak power. Figure 4(b) shows the temperature profiles along the ridge-lines denoted in Fig. 4(a). We see that a thicker GST layer can absorb more energy, resulting in a higher temperature. However, as there is a large gradient in temperature, switching the entire GST cell is more difficult for thick GST, especially in strip waveguide. The complete amorphization is hard to obtain for a longer than 1 μm GST cell. Figure 4(c) shows the maximum temperature change as a function of the GST thickness. With the same thickness of GST, its temperature in the strip waveguide is 2.5-3 times higher than that in the MMI waveguide. In order to get the same degree of amorphization, the optical pump power required by the MMI waveguide is at least 2.5 times higher than that of the strip waveguide.

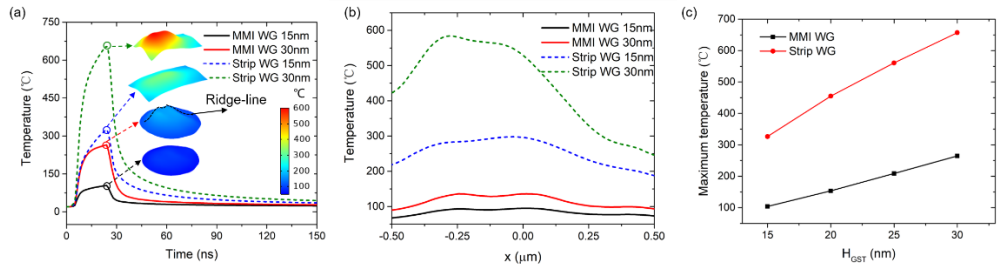

Fig. 4 (a) GST temperature transient response for different thicknesses of GST. Insets: temperature profile in the planar cross-section of the GST layer. (b) Temperature distribution along the ridge-lines. (c) The maximum temperature change as a function of GST thickness.

To verify our simulation predictions, we fabricated an MMI waveguide covered with 30-nm-thick GST on top (Device I) and two strip waveguides covered with 30-nm-thick (Device II) and 15-nm-thick (Device III) GST on top and. The dimensions of the strip waveguide are in agreement with the above simulation. It should be noted that we use MMI crossing waveguide to replace the MMI waveguide in the following test. As shown in Fig. 5(a), GST phase change can be induced by electrical pulses applied by the orthogonal MMI waveguide (see Section 2.3 for details). These three devices were heated for 5 min at 200 °C on a hot plate for implementing a fully crystalline state. Figure 5(b) shows the SEM images of Device II with fully crystalline state GST. We first investigated the amorphization process of the Device I when a rectangular optical pulse was applied. The optical pulse has a fixed duration of 20 ns. Figures 5(c) and 5(d) present the SEM images of the Device I when the peak power of the optical pulse power is 75 mW and 150 mW, respectively. The amorphous area is approximately elliptical for the two cases, which is in agreement with the simulation results. It indicates that a high power pulse can induce more amorphous region. However, as the pulse power is further increased, a larger amorphized zone is hard to be obtained. The temperature at the center of the GST-cell is much higher than its peripheral region, which results in a severe degradation of central part in the GST when its peripheral region reaches the melting point. The Device II (30-nm-thick GST), as explained in the previous part, has a higher attenuation of the optical transmission in both states. If one attempts to reach a higher transmission level in the amorphization process, multiple higher power optical pulses are needed to switch more regions of GST to the amorphous state. However, the GST cell will be irreversibly degraded, as shown in Fig. 5(e). We also experimentally compare the power consumption of the three devices in the amorphization process. Figure 5(f) shows the transmission change as a function of pump pulse power for Device I (green line), Device II (red line), and Device III (black line). The obtained switching contrast of three devices all have a relatively linear response with increased pump pulse power. The amorphization threshold power is higher for a thinner GST cell. We notice that with the same thickness of GST, the MMI waveguide needs more optical power than the strip waveguide to get the same transmission contrast. The power consumption for Device I is 4.4 times higher than Device II, which is higher than expected because the MMI crossing waveguides experience larger thermal dissipation than MMI waveguide. For strip waveguide with different thicknesses of GST, the power consumption in Device III is 2.5 times than that in Device II. This is in good agreement with our simulation results.

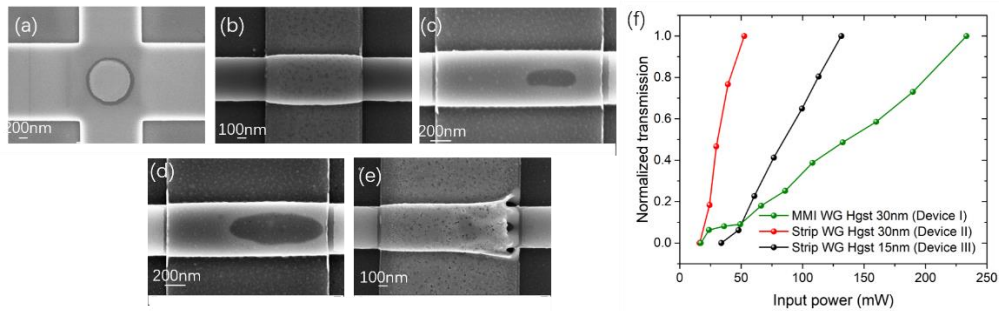

Figure 5 (a, b) SEM images of the Device I (a) and Device II (b) when the GST is in the original fully crystalline state. (c-d) SEM images of Device III when the GST is amorphized by (c) a 75-mW optical pulse, and (d) a 150-mW optical pulse. (e) SEM image of the Device II after multiple high-power optical pulses are applied. (f) Transmission change as a function of optical pump optical pulse power.

*2.3 Dynamic response to electrical pulses*

Although the phase change under optical pulses exhibits a faster response speed, it is difficult to be employed in large-scale optical switch fabrics due to inconvenient routing of the optical pump pulses. Instead, phase change under electrical pulses is more suitable since the electrical pulses can be delivered by electrical wires on another layer. In order to electrically heat up the GST cell, we resort to the crossing structures, as shown in Fig. 6. The center of the crossing is covered by 30-nm-thick GST and 30-nm-thick ITO. The crossed silicon strip is heavily $P^{++}$-doped with a concentration of $10^{20}$ $cm^{-3}$ so that it works as a resistive heater. A Ti/Au (5nm/150nm) layer is deposited on top to provide electrical connection. The gap size between the two Au electrodes is 5 μm. The crossed strips are 1 μm and 1.7 μm wide for the waveguide and MMI crossings, respectively. Figure 6(b) illustrates the optical power propagation in the *x-y* plane for the waveguide and MMI crossings. The crossing induced loss for the regular waveguide and the MMI is 2.34 dB and 0.51 dB, respectively. The lower loss for the MMI crossing is due to its self-imaging principle.

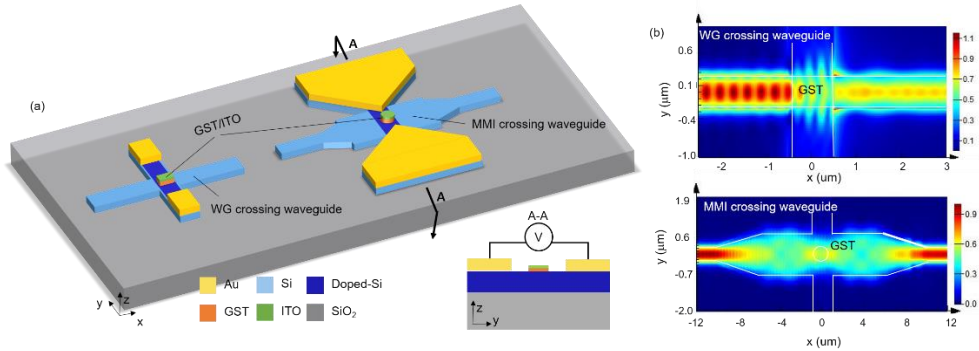

Fig. 6. (a) Strip waveguide crossing (left) and MMI crossing (right) integrated with electrically actuated GST. Inset: cross-section of the GST cell. (b) Simulated electrical field intensity distributions in the strip waveguide crossing (up) and the MMI crossing (bottom).

Next, we compare the power consumption in electrical phase change for the two crossings. Figure 7(a) shows the transient temperature response when an electrical pulse is applied. The pulse duration is 20 ns and the peak voltage is 9 V and 11 V. The insets show the temperature profiles in the lateral cross section of the GST layer. Due to the lower thermal dissipation, the strip waveguide crossing reaches a higher temperature compared with the MMI crossing. Figure 7(b) shows the temperature distributions along the ridge-lines denoted in Fig. 7(a). The temperature variation is less than 100 °C across the GST cell, which indicates that the electrical pulse can induce phase change for the entire GST cell.

Figure 7(c) shows the temperature transient response for different thicknesses of GST in the amorphization process. It reveals that the thickness of GST does not have a significant effect on the temperature response, because the heating process is governed by the Joule heat transfer from the silicon resistive heater. In contrast, for optically induced phase change, the optical pump pulse evanescently interacts with the GST cell, and as a result, a thicker GST cell absorbs more light and reaches a higher temperature.

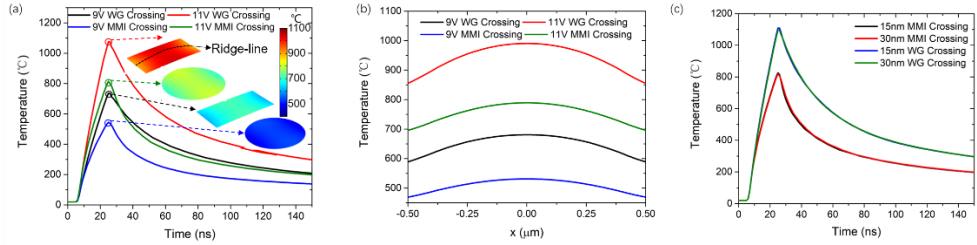

Fig. 7. (a) GST temperature variation in response to an excitation electrical pulse for the strip waveguide crossing and the MMI crossing. Insets: temperature profiles in the planar cross-section of the GST layer. (b) Temperature distribution along the ridge-lines in (a). (c) GST temperature variation for two thicknesses of GST cell.

Table 2 summarizes power consumption, operation speed, insertion loss, and extinction ratio of the two types of devices under optically and electrically induced phase change. The GST thickness is 30 nm. It can be seen that the insertion loss and the extinction ratio of the MMI devices are better than the strip waveguide devices. The power consumption of the MMI is higher than the strip waveguide because of the large silicon area underneath the GST cell. The dead time under electrical pulses is longer than that under optical pulses. Because the doping region beneath the GST is a rectangular strip, heat is uniformly generated along the doping region. Given that the GST is only deposited at the center of the waveguide. The generated heat is only partially utilized to raize the GST temperature. A considerable amount of heat is wasted, leading to larger power consumption. In contrast, for the optically induced phase change, the GST temperature rises through direct light absorbtion, and hence power consuption is lower and dead time shorter.

**Table 2: Performance comparison of the strip waveguide and MMI devices**

| Methods | Devices | IL (dB) | ER (dB) | Power consumption (mW) | Dead time (ns) |
|---|---|---|---|---|---|
| Optial excitation | Strip WG | 0.85 | 11.01 | 53* | 8 |
| | MMI WG | 0.50 | 12.85 | 233* | 6 |
| Electrical excitation | Strip crossing | 2.34 | 7.91 | 324** | 65 |
| | MMI crossing | 0.51 | 12.66 | 484** | 47 |

*optical pulse power; **electrical pulse power

In the MMI crossing, the microheater can be further optimized to ensure heat is concentrated more in the center. In this way, the power consumption can be reduced and the operation speed can be improved. In the following section, we optimize the gap size between the Au electrodes and the shape of the doped region.

### 2.4 Microheater optimization in the MMI crossing

#### 2.4.1 Gap size of the Au electrodes

The separation distance between the two metal electrodes (*gap*) significantly affects the power consumption in the electrical phase transition process. A smaller gap gives a lower resistance of the microheater, which as a result, facilitating the reduction of electrical pulse voltage. Moreover, reducing the gap also makes heat more concentrated in the central part of the MMI crossing. Figure 8(a) shows the normalized transmission change of the MMI crossings at 1550 nm in the amorphization process for three gap sizes of 3.5 μm, 4 μm, and 5 μm. The excitation electrical pulse has a fixed duration of 20 ns and a varying voltage. It can be seen that the pulse voltage for the 5-μm gap saturates at about 11.6 V, while for the 3.5-μm gap, it saturates at 9 V. Therefore, it is feasible to reduce the pulse voltage by using a smaller gap. Figure 8(b) compares the normalized transmission changes of the devices with three gap sizes during the amorphization process. We can observe that the gap size has no influence on the operation speed of the device, but it will affect the power consumption of the device.

Smaller gap size can help reduce power consumption. The simulation results show that the 3-μm gap does not suffer additional optical loss and the driving voltage can be less than 9 V.

*2.4.2 Shape of the doping region*

The doping shape also affects microheater resistance and heat distribution. We investigate three devices with different doping shapes. Device I (T1) is hourglass-shaped and the waist size is smaller 0.25 μm than the diameter of GST cell. Device II (T2) also has an hourglass-shaped doping region, but the waist is larger 0.25 μm than the diameter of the GST disk. The width of both ends is 1.7 μm for Device I and Device II. Device III (T3) has a regular rectangular doping strip. The doping width is equal to the diameter of GST cell. Figures 8(c) and 8(d) show the normalized transmission change as a function of pulse width (Fig. 8(c)) and pulse amplitude (Fig. 8(d)) for the crystallization process. Changing the doping region from a rectangular strip to an hourglass shape can reduce the pulse voltage because heat is mostly generated and localized in the narrow waist part. For the electrical pulse with the same amplitude and width, a higher temperature is obtained in the GST area for Device I, leading to low power consumption in electrical phase change.

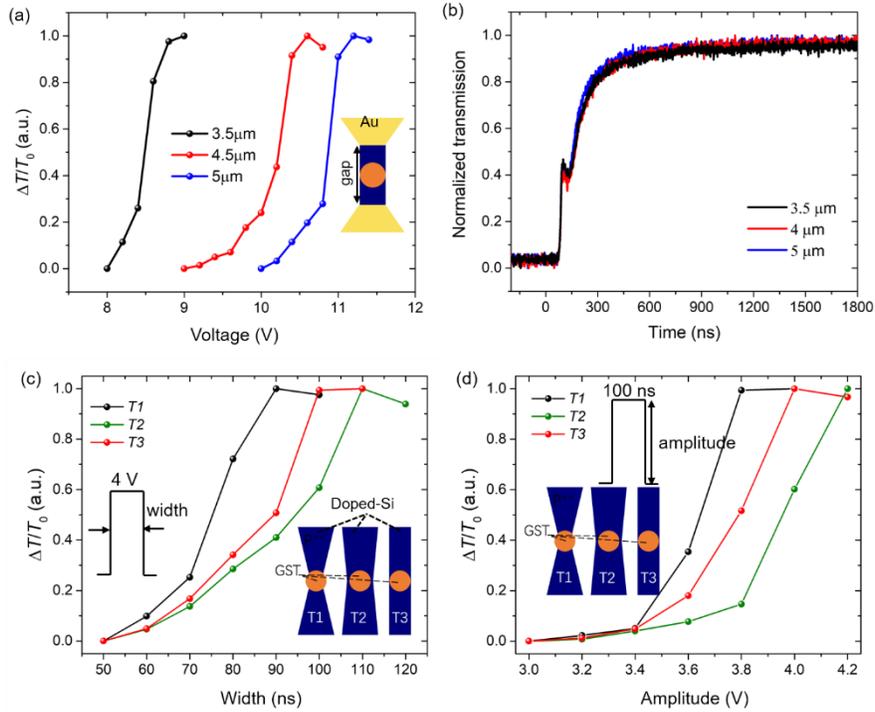

Fig. 8. (a) Normalized transmission change varies with the electrical pulse amplitude for three gap sizes. (b) Normalized temporal response for the device with three gap sizes in the amorphous process. (c, d) Normalized transmission change with (c) electrical pulse width and (d) electrical pulse amplitude for three doping shapes in the crystallizaiton process.

## 3. Conclusion

In conclusion, we used two types of Si-GST hybrid waveguides to investigated the differences in the reversible phase transitions induced by optical and electrical pulses. For the phase transition induced by optical pulses, the strip waveguide has a clear advantage in terms of power consumption due to lower thermal dissipation. The MMI waveguide is feasible to achieve a larger extinction ratio and a higher operation speed due to the larger volume and thermal conductivity of silicon material. However, all-optical switching has difficulty in switching large-area GST and need sophisticated light routing systems. The disadvantage can be overcome by adding crossing structures to apply electrical pulses. Although the crossing does not induce additional loss for the MMI, it increases the power consumption and dead time. By optimizing the gap size between the two Au electrodes and the shape of the doped region, we experimentally demonstrate that a smaller gap size and hourglass-shaped doping region can decrease the power consumption. These devices offer a promising solution for specific applications like large-scale rewritable optical circuits with a small footprint and high energy efficiency. The method built in our work also assists the design and optimization of the Si-GST hybrid integration devices such as nonvolatile optical switches, optical memories, and reconfigurable photonic circuits.

## 4. Acknowledgments

The authors would like to thank the support from the National Natural Science Foundation of China (NSFC) (61705129, 61535006), the National Key R&D Program of China (2018YFB2201702), and the Shanghai Municipal Science and Technology Major Project (2017SHZDZX03).